\documentclass[runningheads]{llncs}
\usepackage{url}
\usepackage{color}
\usepackage{graphicx}
\usepackage{amsmath}
\usepackage{float}
\usepackage{url}
\usepackage{hyperref}

\usepackage{amssymb}
\usepackage{comment}
\usepackage{xspace,framed}
\usepackage[ruled,vlined]{algorithm2e}
\usepackage{algorithmic}
\usepackage{soul}
\usepackage{times}

\usepackage[dvipsnames]{xcolor}

\def\orcidID#1{\smash{\href{http://orcid.org/#1}{\protect\raisebox{-1.25pt}{\protect\includegraphics{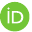}}}}}

\newcommand{\fb}{\textit{FuSeBMC}}
\newcommand{\tracer}{\textit{Tracer}}

\newcommand{\cou}[1]{\texttt{#1}}

\begin{document}

\title{FuSeBMC v4: Smart Seed Generation for Hybrid Fuzzing}
%\subtitle{(Competition Contribution)}

\author{Kaled M. Alshmrany\inst{1,2}\orcidID{0000-0002-5822-5435} \and Mohannad Aldughaim\inst{1}\orcidID{0000-0003-1708-1399} \and Ahmed Bhayat\inst{1}\orcidID{0000-0002-1343-5084} \and \\ Lucas C. Cordeiro\inst{1}\orcidID{0000-0002-6235-4272} }

\institute{
    University of Manchester, Manchester, UK \and Institute of Public Administration, Jeddah, Saudi Arabia
}

\authorrunning{K. M. Alshmrany et al.}

\titlerunning{FuSeBMC v.4: Smart Seed Generation for Hybrid Fuzzing}

\maketitle  
\vspace{-4mm}
\begin{abstract}
\fb{} is a test generator for finding security vulnerabilities in C programs. In earlier work \cite{alshmrany2021fusebmc}, we described a previous version that incrementally injected labels to guide Bounded Model Checking (BMC) and Evolutionary Fuzzing engines to produce test cases for code coverage and bug finding. This paper introduces a new version of \fb{} that utilizes both engines to produce smart seeds. First, the engines are run with a short time limit on a lightly instrumented version of the program to produce the seeds. The BMC engine is particularly useful in producing seeds that can pass through complex mathematical guards. Then, \fb~runs its engines with more extended time limits using the smart seeds created in the previous round. \fb{} manages this process in two main ways using its \tracer{} subsystem. Firstly, it uses \emph{shared memory} to record the labels covered by each test case. Secondly, it evaluates test cases, and those of high impact are turned into seeds for subsequent test fuzzing. As a result, we significantly increased our code coverage score from last year, outperforming all tools that participated in this year's competition in every single category.
\end{abstract}

%====================================================================
\vspace{-5.9mm}
\section{Overview}
\label{sec:overview}
\vspace{-1mm}
%====================================================================

Software testing is one of the most crucial phases in software development ~\cite{kosindrdecha2010test}. Tests often expose critical bugs in software applications. In earlier work~\cite{alshmrany2021fusebmc}, we presented \fb{}, an automated test generation tool that exploits the combination of Fuzzing and Bounded Model Checking. \fb{} achieved second place in Test-Comp 2021~\cite{beyer12649status,alshmrany2020fusebmc} and first place in the \textit{Cover-Error} category. It ranked fourth in the \textit{Cover-Branches} category. This year, we introduce a new version of \fb{} (v4) that adds smart seed generation and shared memory amongst other improvements and features. The new version significantly improves on the previous version, particularly relating to code coverage. One of the primary contributions of this paper is the linking of a grey-box fuzzer with a bounded model checker. A bounded model checker works by treating a program as a state transition system and then checking whether there exists a transition in this system of length less than a bound $k$ that violates the property to be verified \cite{Biere09,CordeiroFM12}. We leverage this power of model checkers as a method for smart seed generation. During grey-box fuzzing, if a particular branch has not been explored, BMC can be used to provide a model (set of assignments to input variables), which reaches the branch. This model is then shared in memory and seeded for further grey-box fuzzing. However, BMC can be slow and resource-intensive. We also carry out a lightweight static program analysis to recognize input verification. We analyze the code for conditions on the input variables and ensure that seeds are only selected if they pass these conditions. Together, these contributions turn \fb{} into a world-leading fuzzer.

%====================================================================
\vspace{-2mm}
\section{Test Generation Approach}
\label{sec:approach}
\vspace{-1.9mm}
%====================================================================

\begin{figure}
    \includegraphics[width=12cm, height=5cm]{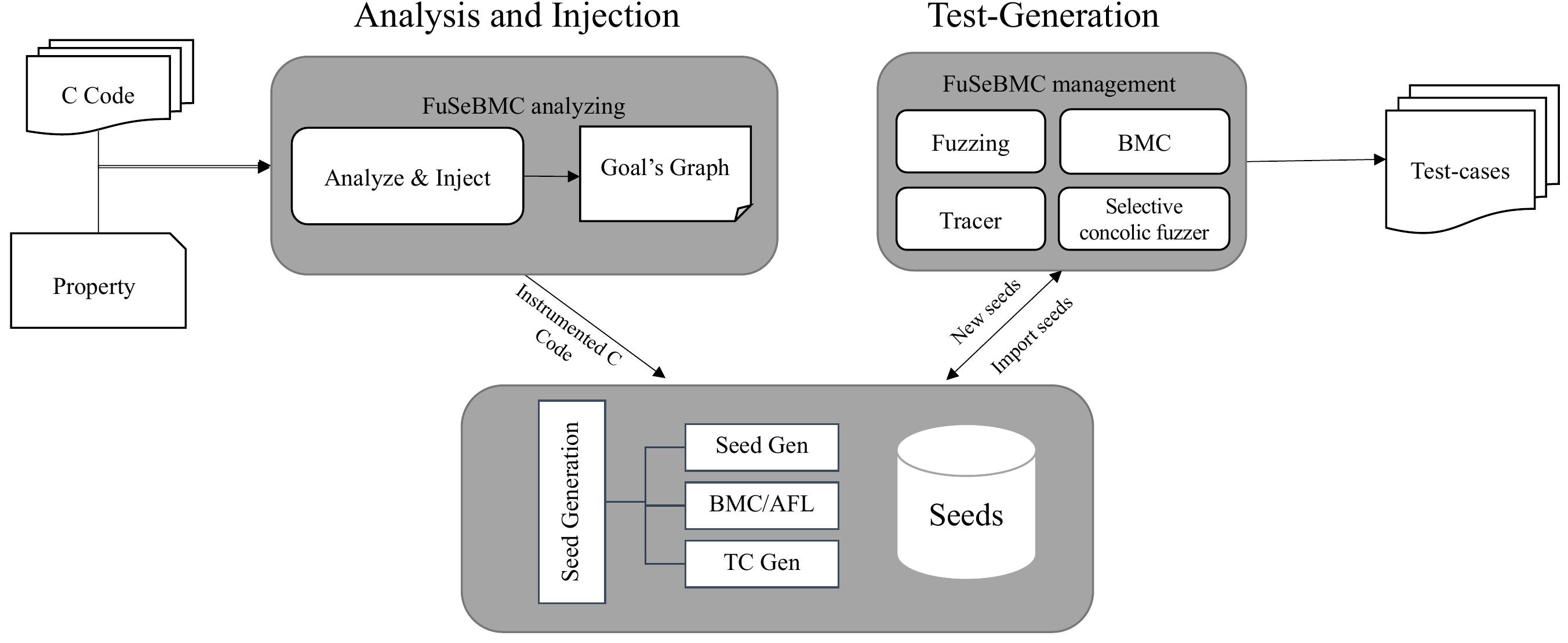}
    \caption{\textit{FuSeBMC} v4 Framework. This figure displays the major components of the \fb{} test generator and how they interact. Note in particular the seed store.}
    \label{fig:framework}
\end{figure}

Figure \ref{fig:framework} provides an overview of the components within \fb{} and how these interact. \fb{} makes use of the Clang tooling infrastructure~\cite{CLANG} to instrument programs. In addition, \fb{} employs three engines in its reachability analysis. \fb{} starts by injecting goal labels into the given input (C program) and ranks them depending on the chosen strategy. After that, \fb{} employs two fuzzers, one based on the American Fuzzy Loop (AFL)~\cite{bohme2017directed,americanfuzzylop_2021}, and a second custom fuzzer, which we refer to as \emph{selective fuzzer} (see~\cite{alshmrany2021fusebmc} for details). ESBMC~\cite{gadelha2019esbmc,gadelha2020esbmc} is a state-of-the-art SMT-based bounded model checker which \fb{} utilises to produce seeds and test cases. Besides for these engines, \fb{} also incorporates a subsystem we call the \tracer{}. This component maintains a program graph and records which labels in the graph have been covered by which test cases. It coordinates the other tools and handles the passing of information between them. The fuzzers generate test cases by randomly mutating the program's input and running it to analyze code coverage. The BMC engine executes the given program symbolically to determine the reachability of particular goal labels. In case of success, a witness -- a set of inputs leading to the goal --  is produced. If any engine manages to reach a label, we say that label is \emph{covered}. The \tracer{} records which goals have been covered by which test cases. This information is used to prevent the computationally expensive BMC engine from trying to reach an already covered goal. More importantly, it is used to coordinate the fuzzing and BMC engines. Let $L_n$ be the deepest label on some branch that ESBMC covers and let $L_{n+1}$ be the next deepest label on the branch. \tracer{} records the test case produced by ESBMC that covers $L_n$ and then passes this as a seed to the fuzzer. Mutating such a seed has a far larger chance of leading to a test case that covers $L_{n+1}$ than starting the fuzzer from scratch. \fb{} runs until all goals are covered or a timeout is reached.

\vspace{-2mm}
\vspace{-2mm}

%-----------------------------------------------------
\vspace{-2mm}
\subsubsection{Code Instrumentation}
%-----------------------------------------------------

\fb{} front-end uses Clang tooling infrastructure~\cite{CLANG} to parse a C program and produce an Abstract Syntax Tree (AST). While traversing the AST, \fb{} injects labels into each branch, including every conditional statement, loop, and function. Using these labels, \fb{} can measure the code coverage. 

%-----------------------------------------------------
\vspace{-2mm}
\subsubsection{Reachability Graph Analysis}
%-----------------------------------------------------

After instrumenting the C program, \fb{} analyzes it and produces a reachability graph. The graph will assign each goal label to the code block it is located in. Then, \fb{} ranks goals depending on the strategy chosen. For example, one strategy is to prefer deeper goals over shallower goals. This strategy improves the performance of \fb{} since a test case that covers a deep goal will also cover shallower goals on the path to it. \fb{} also ranks coverage metrics over others, such as conditional coverage over loop coverage. 

%-----------------------------------------------------
\vspace{-2mm}
\subsubsection{Seed Generation}
%-----------------------------------------------------

A unique aspect of the latest version of \fb{} is a seed generation phase that is run prior to the start of the principal reachability analysis. In this phase, \fb{} first lightly instruments the code under test by limiting loop bounds and assuming a narrow range of values for input variables. The bounds on input variables are further limited by carrying out a lightweight static analysis to recognize code that applies verification conditions to input variables. After instrumenting the code, \fb{} runs its fuzzing and BMC engines with very short time limits. The test cases generated by these engines are ranked, and the highest impact test cases are selected as smart seeds for the next round. The impact of a test case is measured using two metrics. First, the number of labels covered uniquely by that test case, and second, the program depth achieved by the test case. ESBMC is particularly effective at seed generation as its underlying SMT solvers can be used to discover test cases that circumvent complex mathematical guards. 

%-----------------------------------------------------
\vspace{-2mm}
\subsubsection{Reachability Analysis Engines}
%-----------------------------------------------------

In its primary phase, \fb{} carries out reachability analysis.  Essentially this involves running the engines with longer timeouts on the original non-instrumented code with the fuzzer making use of the smart seeds. The \tracer{} coordinates the various engines through the use of \emph{shared memory}. For example, assume that ESBMC is unable to cover some goal $L$ and let $L'$ be the deepest goal on the path to $L$ that ESBMC \emph{can} cover. The tracer records the test case that covers $L'$ and passes it to the fuzzer via shared memory as an incomplete seed. Thus, \fb{} combines the strengths of both types of engines. The BMC engine produces seeds that bypass complex guards and thereby help the fuzzers explore paths deep within the program. During reachability analysis, the \tracer{} constantly monitors the test cases produced by the various engines. High impact test cases, as measured by the metrics discussed above, are passed by the \tracer{} back to the seed store. Thus, the seed store is dynamically updated as the analysis progresses.

%\vspace{-2mm}
%\begin{figure}
%\centering
%    \includegraphics[width=12cm, height=3.5cm]{Tracer.pdf}
%    \caption{\textit{FuSeBMC} v4 - \tracer. \textcolor{red}{Add some short explanation about this figure here.}}
%    \label{fig:tracer}
%\end{figure}

%====================================================================
\section{Strengths and Weaknesses}
\label{sec:strengths-weaknesses}
\vspace{-2mm}
%====================================================================
 The strengths of the latest version of \fb{} are as follows. It runs a dedicated seed generation phase in order to start the main fuzzing effort with high quality, high impact seeds. Furthermore, during the main test-generation phase, these seeds are constantly being updated. Beyond this, it incorporates a dedicated subsystem, the \tracer{}, that uses a shared memory store to manage the various engines. By combining the engines, the \tracer{} ensures that \fb{} far outperforms the individual engines, or even the running of the engines in parallel, but isolated. The outcome of these improvements can be seen in the ECA and Combination benchmark sets.  Previously, these posed a challenge to \fb{}. With the latest changes, according to preliminary results, \fb{} achieved first place in the Combination subcategory and took second place in the ECA subcategory of the 2022 Test-Comp competition. In the ECA subcategory \fb{} displayed a remarkable $60$\% improvement over its performance from the previous competition. The preliminary results also indicate that \fb{} has achieved first place in the \textit{Cover-Branches} category with high coverage and validation statistics. One of the weaknesses of \fb{} that we plan to work on, is that for large programs, particularly for programs that redefine C library functions, seed generation can be slow and consume too much of the tool's time.
%Overall, in both \textit{Cover-Error} and \textit{Cover-Branches} categories, various test cases produced by \fb{} are validated successfully. For instance, in the \textit{Cover-Error} category, TestCov confirms $731$ test cases produced by \fb{} out of $777$ with a percentage reach of $94$\%. While in \textit{Cover-Branches}, \fb{} reached $61$\% of coverage. Moreover, our result in \textit{Cover-Branches} has improved by $16$\% compared to the previous version and placed \fb{} in first place in this category.

%====================================================================
\vspace{-2mm}
\section{Tool Setup and Configuration}
\vspace{-2mm}
%====================================================================

\fb{} can be run using the command below. The user is required to set the architecture, the property file path, the competition strategy, and the benchmark path, as:

\vspace{.5em}
\texttt{fusebmc.py [-a \{32, 64\}] [-p PROPERTY\_FILE]\newline \hphantom{this text is inv} [-s \{kinduction,falsi,incr,fixed\}]\newline \hphantom{this text is inv} [BENCHMARK\_PATH]}
\vspace{.5em}

\noindent where \cou{-a} sets the architecture to 32 or 64, \cou{-p} sets the property file to \cou{PROPERTY\_FILE}, where it has a list of all the properties to be tested. \cou{-s} sets the BMC strategy to one of the listed strategies\cou{\{kinduction,falsi,incr,fixed\}}. The Benchexec tool info module is \cou{fusebmc.py} and the benchmark definition file is \cou{FuSeBMC.xml}.

%====================================================================
\vspace{-2mm}
\section{Software Project}
\vspace{-2mm}
%====================================================================

\fb{} is implemented using C++, and it is publicly available under the terms of the MIT License at GitHub\footnote{\url{https://github.com/kaled-alshmrany/FuSeBMC}}. The repository includes the latest version of \fb{} (version 4.1.14). \fb{} dependencies and instructions for building from source code are all listed in the \cou{README.md} file.

%====================================================================
\bibliographystyle{plainnat.bst}
%\bibliography{../refs}
\bibliography{References.bib}
%====================================================================

%\begin{thebibliography}{1}

%\end{thebibliography}
%\subsubsection{Open Access} This chapter is licensed under the terms of the Creative Commons Attribution 4.0 International License (\url{http://creativecommons.org/licenses/by/4.0/}), which permits use, sharing, adaptation, distribution and reproduction in any medium or format, as long as you give appropriate credit to the original author(s) and the source, provide a link to the Creative Commons license and indicate if changes were made.

%The images or other third party material in this chapter are included in the chapter’s Creative Commons license, unless indicated otherwise in a credit line to the material. If material is not included in the chapter’s Creative Commons license and your intended use is not permitted by statutory regulation or exceeds the permitted use, you will need to obtain permission directly from the copyright holder.
\end{document}